# Benchmarking Federated Learning in Edge Computing Environments: A Systematic Review and Performance Evaluation

Sales G. Aribe Jr. * and Gil Nicholas T. Cagande

Information Technology Department, Bukidnon State University, Malaybalay City, Philippines
Email: sg.aribe@buksu.edu.ph (S.G.A.J.); gilcagande@buksu.edu.ph (G.N.T.C.)
*Corresponding author

***Abstract*—Federated Learning (FL) has emerged as a transformative approach for distributed machine learning, particularly in edge computing environments where data privacy, low latency, and bandwidth efficiency are critical. This paper presents a systematic review and performance evaluation of FL techniques tailored for edge computing. It categorizes state-of-the-art methods into four dimensions: optimization strategies, communication efficiency, privacy-preserving mechanisms, and system architecture. Using benchmarking datasets such as MNIST, CIFAR-10, FEMNIST, and Shakespeare, it assesses five leading FL algorithms across key performance metrics including accuracy, convergence time, communication overhead, energy consumption, and robustness to non-Independent and Identically Distributed (IID) data. Results indicate that SCAFFOLD achieves the highest accuracy (0.90) and robustness, while Federated Averaging (FedAvg) excels in communication and energy efficiency. Visual insights are provided by a taxonomy diagram, dataset distribution chart, and a performance matrix. Problems including data heterogeneity, energy limitations, and repeatability still exist despite advancements. To enable the creation of more robust and scalable FL systems for edge-based intelligence, this analysis identifies existing gaps and provides an organized research agenda in the future.**

## I. Introduction

The amount of data generated at the network edge has skyrocketed due to the quick spread of smart devices, sensors, and Internet of Things (IoT) technologies. The increasing needs for real-time analytics, low-latency processing, and data privacy are difficult for traditional cloud-centric architectures to handle, particularly in latency-sensitive applications like industrial automation, remote healthcare, and driverless cars [1]. In order to overcome these obstacles, edge computing has become a distributed computing paradigm that improves responsiveness and decreases reliance on centralized servers by bringing processing and data storage closer to data sources [2].

Meanwhile, Federated Learning (FL) has been presented as a viable approach to machine learning in distributed environments while maintaining privacy [3]. FL protects user privacy and reduces communication overhead by allowing several dispersed devices or nodes to work together to train a common model without sharing raw data [4]. Because of this feature, FL is a perfect partner for edge computing, where privacy is crucial and data is naturally decentralized.

Notwithstanding the expanding corpus of research on FL, there are still obstacles to overcome to successfully integrate it into edge computing environments. These include the statistical heterogeneity of local data, commonly referred to as non-Independent and Identically Distributed (IID), along with unstable network conditions, computational resource limitations, and energy constraints [5, 6]. To solve these problems, a wide variety of FL algorithms have been put forth, ranging from robust optimization strategies and privacy-enhancing protocols to communication-efficient approaches such as quantization and compression. However, the field currently lacks a comprehensive review that not only categorizes these techniques but also systematically benchmarks them against practical performance metrics relevant to edge scenarios.

This review addresses that gap by offering a systematic and comparative analysis of federated learning techniques designed for edge computing contexts. Specifically, it aims to:

- classify FL approaches based on their optimization strategies, communication models, privacy mechanisms, and system architecture;
- benchmark selected algorithms using a performance matrix that evaluates accuracy, convergence time, communication overhead, energy efficiency, and privacy robustness;
- identify technical challenges and research gaps that remain unresolved; and

- provide recommendations and future directions to guide the development of scalable and efficient FL systems for edge environments.

By offering a structured evaluation framework and consolidating current trends, this review serves as a foundational reference for researchers, practitioners, and system designers seeking to implement and optimize FL in edge computing settings.

## II. Background and Theoretical Foundations

### A. Edge Computing Overview

Edge computing is a decentralized computing paradigm that processes data at or near the source of data generation rather than relying solely on centralized cloud servers. It involves distributing computing resources to edge devices such as sensors, gateways, and local servers, thereby reducing latency and bandwidth usage while improving response times [1]. Architecturally, edge computing extends the cloud toward the user by deploying mini data centers or computational nodes closer to end-users, forming a hierarchical structure comprising the cloud, fog, and edge layers [7].

One of the key advantages of edge computing is its ability to support latency-sensitive applications by minimizing the distance between data source and processing unit. This enables real-time analytics and decision-making in domains such as autonomous vehicles, remote surgery, and industrial automation [2]. Moreover, by processing data locally, edge computing reduces data transfer volumes, thus alleviating bandwidth bottlenecks and mitigating privacy risks, which are especially critical under regulatory frameworks like General Data Protection Regulation.

Throughout this review, several operational concepts are used to characterize federated learning performance. "Communication overhead" refers to the amount of data transmitted between clients and servers (or peers) per communication round, typically measured in megabytes and influenced by model size, compression ratio, and upload/download frequency. "Heterogeneity" encompasses three dimensions commonly encountered in federated settings: (1) statistical heterogeneity, where client data are non-IID due to differences in user behavior or contextual factors; (2) system heterogeneity, which reflects variation in client hardware capabilities such as CPU architecture, memory, and available energy; and (3) network heterogeneity, arising from fluctuating bandwidth, latency, and intermittent connectivity in edge and mobile environments. These clarifications provide a quantitative and multidimensional basis for understanding challenges highlighted in the subsequent sections.

### B. FL Fundamentals

FL is a decentralized machine learning approach where multiple clients, typically mobile or edge devices, collaboratively train a shared global model without sharing raw data [4]. Instead, clients compute model updates locally and send only the encrypted gradients or model parameters to a central aggregator, which updates the global model. This design promotes data privacy and enables learning from decentralized, sensitive, or proprietary datasets.

While FL avoids the transmission of raw data, which reduces privacy risks and prevents the large one-time data transfers typical of centralized training, it does not automatically reduce overall communication requirements. In practice, FL often increases the frequency of communication because edge devices must repeatedly exchange model updates with aggregators across multiple training rounds. This iterative communication pattern can lead to substantial communication overhead, particularly in bandwidth-constrained edge environments. As later discussed in Section V and highlighted by recent studies, communication remains one of the most significant bottlenecks in FL despite its advantages in privacy preservation.

The standard FL workflow involves: (1) initializing a global model on a central server; (2) broadcasting the model to selected edge clients; (3) clients training the model on local data; and (4) sending updates to the server for aggregation, typically using Federated Averaging (FedAvg) [5]. Fig. 1 illustrates the typical architecture of a federated learning system in an edge computing environment, including the role of edge devices, fog nodes, and the cloud server in the training and aggregation process.

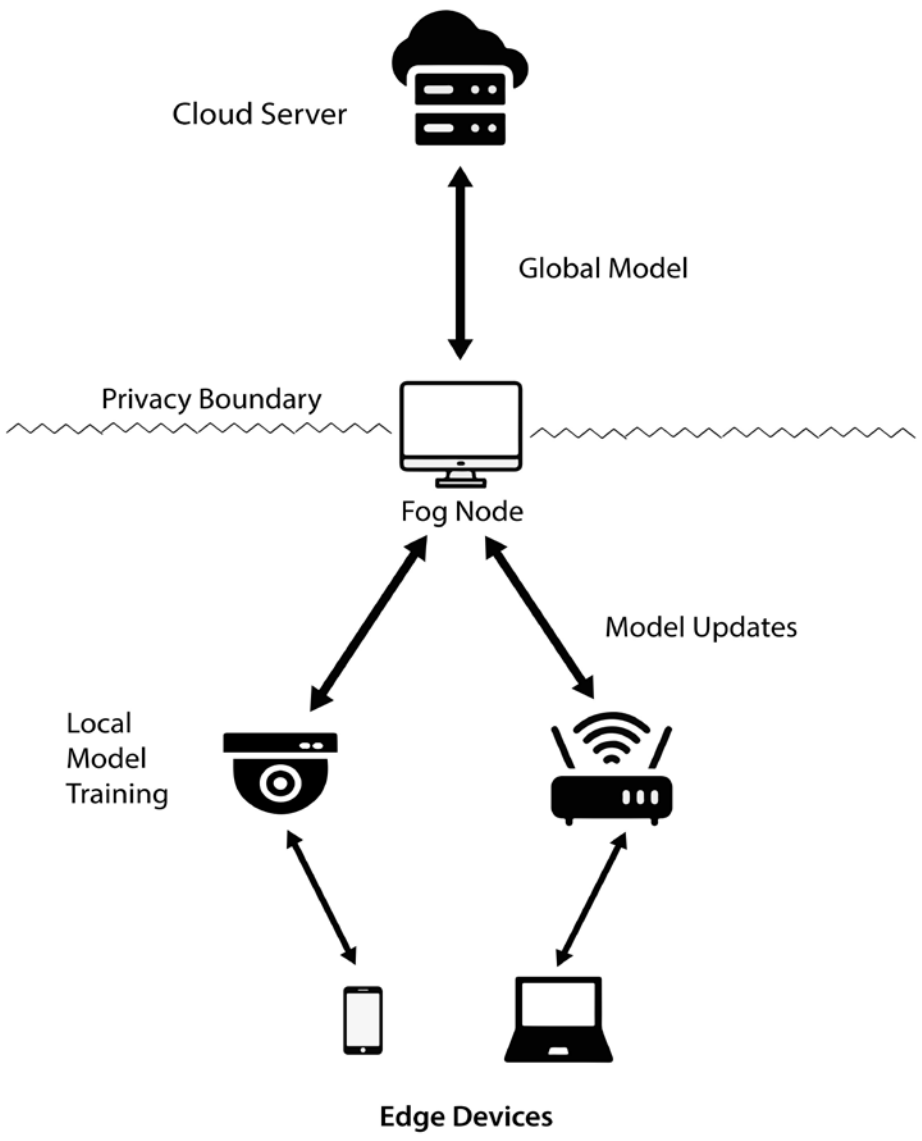

Fig. 1. Federated Learning (FL) workflow in edge computing.

Edge devices (e.g., mobile phones, sensors) perform local model training and send encrypted updates to a fog node aggregator. The fog node collects and forwards aggregated updates to the cloud server, which sends back the improved global model. A privacy boundary is maintained, ensuring that raw data never leaves the local devices.

FL can be categorized into three main types based on the distribution of data across clients [8]:

- Horizontal FL: Clients share the same feature space but differ in data samples (e.g., banks with similar data structures but different customers);

- Vertical FL: Clients share the same data samples but have different feature spaces (e.g., a hospital and insurance company serving the same patients but recording different attributes);
- Federated Transfer Learning: Both the sample space and feature space differ across clients, and knowledge transfer is used to bridge differences.

These variations enable FL to adapt to diverse collaborative environments while safeguarding data locality and confidentiality.

## III. Methodology of the Review

This section outlines the structured methodology adopted to ensure the transparency, rigor, and reproducibility of this review. The review follows guidelines inspired by the Preferred Reporting Items for Systematic Reviews and Meta-Analyses (PRISMA) framework [9] and the Search, Appraisal, Synthesis, and Analysis (SALSA) [10]. The objective was to identify and synthesize studies that investigate the implementation and performance of FL in edge computing environments.

### *A. Review Protocol*

To ensure the relevance and quality of included studies, a set of inclusion and exclusion criteria was developed. The inclusion criteria for eligible articles were as follows: (1) peer-reviewed journal or conference papers, (2) written in English, (3) published between January 2017 and June 2025, and (4) containing original experimental data focused on FL applied within edge computing contexts. Studies were required to report quantitative results using relevant performance metrics such as accuracy, communication cost, convergence time, or energy efficiency. Articles were excluded if they were purely theoretical without empirical data, review papers lacking benchmarking content, or non-peer-reviewed sources such as opinion pieces, workshop abstracts, or editorials.

A comprehensive search was conducted using major electronic academic databases including IEEE Xplore, Scopus, SpringerLink, ScienceDirect, ACM Digital Library, and arXiv. Search queries combined Boolean operators and keywords such as “federated learning” AND “edge computing,” “optimization” “OR communication” OR “privacy” OR “system architecture”, “non-IID” OR “data heterogeneity” OR “client reliability” OR “system scalability” OR “energy efficiency” OR “security” OR “benchmarking” and “accuracy” OR “convergence” OR “energy consumption” OR “non-IID robustness” OR “privacy mechanism” OR “communication overhead”. The initial search yielded 602 articles. After removing duplicates and applying the inclusion and exclusion criteria, 308 articles were retained for full-text review and data extraction. These articles represent the most relevant and empirically grounded studies in the intersection of FL and edge computing.

### *B. Data Extraction and Analysis*

For each selected study, a standardized data extraction template was used to collect critical information including the FL algorithm used, datasets applied (e.g., MNIST, CIFAR-10, FEMNIST, Shakespeare), system deployment environment (e.g., simulated edge platforms or real-world devices), and the set of performance metrics reported. The extracted studies were then classified according to four primary dimensions: (1) optimization strategy, such as FedAvg, FedProx, or SCAFFOLD; (2) communication model, including synchronous, asynchronous, and model compression techniques; (3) system architecture, such as client-server, hierarchical, or peer-to-peer configurations; and (4) privacy mechanisms, including differential privacy, homomorphic encryption, and secure multiparty computation.

The review also synthesized and benchmarked the selected algorithms using standardized performance indicators. Key metrics considered were model accuracy, convergence time (measured in rounds or epochs), communication overhead (typically in megabytes transferred per round), energy consumption (in watts or estimated device power usage), robustness to non-IID data, and the level of privacy guarantees offered. When necessary, reported performance values were normalized to facilitate meaningful cross-study comparisons. The benchmarking results were compiled into a comparative matrix that highlights the relative strengths, weaknesses, and trade-offs of each approach within edge environments. This comprehensive synthesis provides both a theoretical and empirical foundation for identifying promising FL strategies and informing future deployments at the edge.

To further visualize the research trends and thematic concentration of the reviewed studies, a word cloud of author keywords was generated using the bibliometrix package in R, based on data imported from Scopus as shown in Fig. 2.

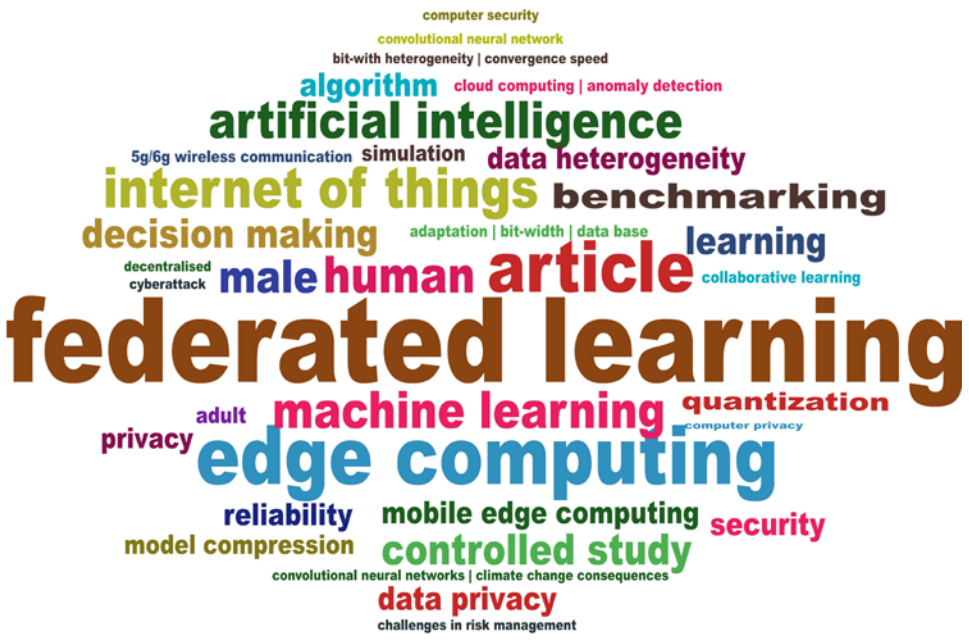

Fig. 2. Word cloud of author keywords from the reviewed articles (2017–2025).

## IV. Taxonomy of Federated Learning Techniques for Edge Computing

To understand how FL methods are adapted to edge computing environments, it is important to classify them according to their core design principles. This section presents a four-dimensional taxonomy based on (1) optimization techniques, (2) communication efficiency, (3) privacy enhancements, and (4) system architectures. This classification enables a systematic evaluation of how various algorithms address the unique constraints of edge environments, including limited bandwidth, computational power, and data heterogeneity.

Fig. 3 provides a visual taxonomy of these FL techniques, summarizing the key methods under each category. This classification supports a structured understanding of the landscape of FL research and facilitates comparative evaluation across approaches.

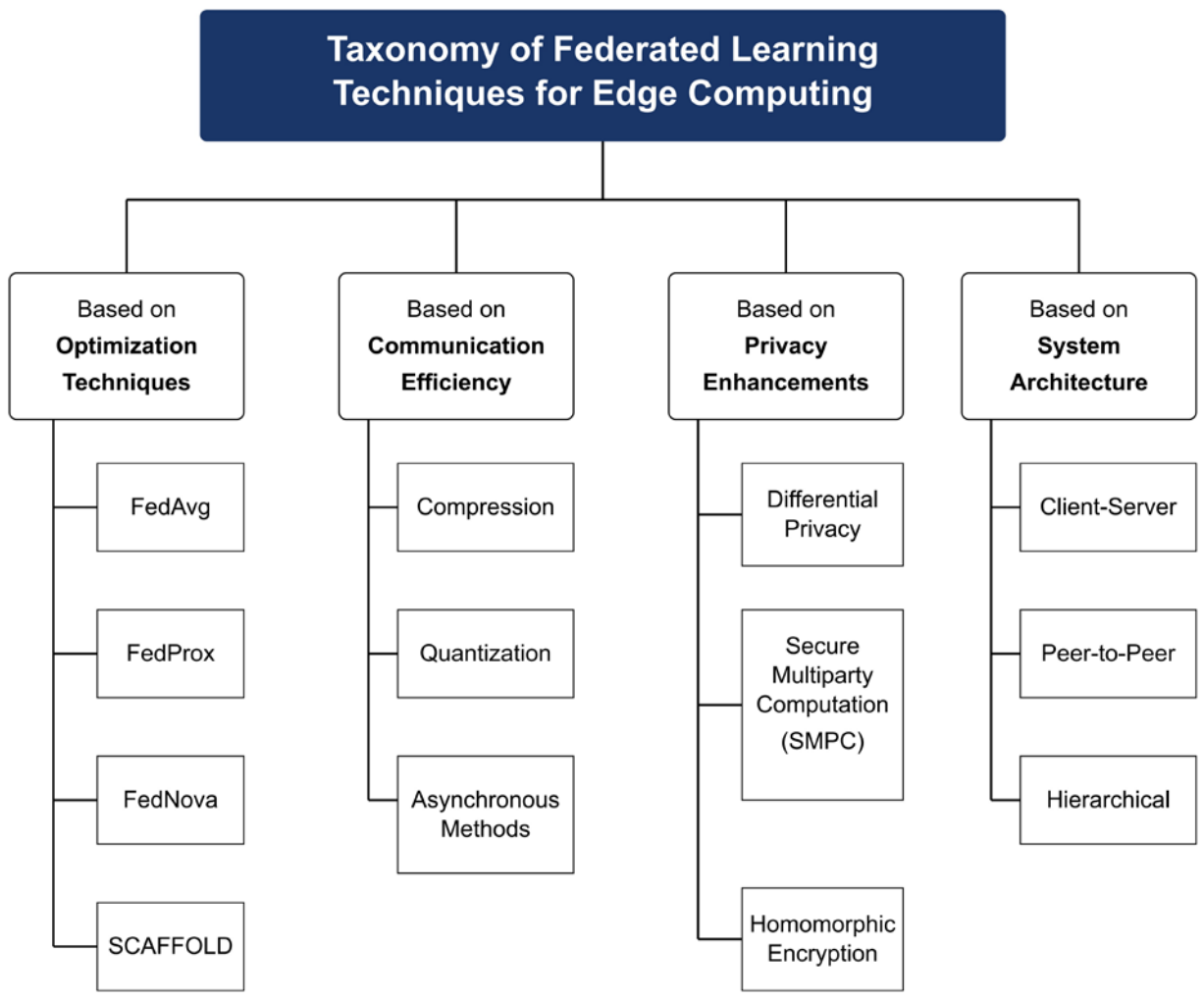

Fig. 3. Taxonomy of FL techniques for edge computing.

The diagram classifies FL methods into four primary categories: (1) Optimization Techniques (e.g., FedAvg, FedProx), (2) Communication Efficiency (e.g., Compression, Asynchronous Methods), (3) Privacy Enhancements (e.g., Differential Privacy, Secure Multiparty Computation), and (4) System Architectures (e.g., Client-Server, Peer-to-Peer, Hierarchical). This taxonomy reflects the diverse strategies used to adapt FL to the resource-constrained, distributed nature of edge computing environments.

### *A. Based on Optimization Techniques*

Optimization lies at the heart of FL algorithm design, particularly in edge settings where data is often non-IID and devices vary in computational capabilities. The most fundamental algorithm is Federated Averaging (FedAvg), introduced by McMahan [4], which averages locally computed gradients or weights across selected clients after each communication round. While simple and effective under IID conditions, FedAvg's performance degrades significantly when data is non-IID.

To address this, FedProx introduces a proximal term in the local objective function to limit the divergence of local updates from the global model, improving convergence under heterogeneous data distributions [6]. FedNova further enhances fairness and convergence by normalizing update contributions based on local step sizes, thereby mitigating client imbalance [11]. Meanwhile, SCAFFOLD employs control variates to correct for client-drift induced by non-IID data, achieving faster convergence and better accuracy [12].

These optimization-oriented techniques aim to improve model generalizability, accelerate training, and reduce sensitivity to data heterogeneity, key concerns in real-world edge deployments.

### *B. Based on Communication Efficiency*

Communication remains one of the most critical bottlenecks in FL, especially in bandwidth-constrained edge environments. To reduce communication cost, several techniques have been proposed. Model compression and quantization are widely used to reduce the size of transmitted updates. For instance, in the study of Konečný *et al.* [13], techniques like sparsification, ternarization, and low-bit quantization transmit only a subset of significant gradient updates.

Asynchronous communication is another strategy, where clients transmit updates at different times rather than in synchronized rounds, thus reducing idle time and improving training throughput [14]. Other adaptive schemes dynamically select clients based on their availability or network conditions to minimize redundant communication and straggler effects [15].

These communication-efficient approaches are essential for scalable FL across thousands of heterogeneous edge devices with fluctuating network connectivity.

### *C. Based on Privacy Enhancements*

While FL is inherently privacy-preserving by design, since raw data is never centralized, it is still vulnerable to indirect attacks such as gradient inversion and membership inference. To enhance privacy guarantees, researchers have integrated advanced cryptographic and differential privacy techniques.

Differential Privacy (DP) adds calibrated noise to local updates or global aggregations, offering formal guarantees against re-identification of individual data points [16]. Bonawitz *et al.* [17] used Secure Multiparty Computation (SMPC) to enable multiple parties to jointly compute functions (e.g., model updates) without revealing their individual inputs, protecting data during transmission and aggregation. Homomorphic Encryption, though computationally intensive, allows operations to be performed on encrypted data, preserving privacy even during processing [18].

Each of these techniques balances trade-offs between security, computational overhead, and model utility, an essential consideration for real-world applications in healthcare, finance, and smart cities.

### *D. Based on System Architectures*

The architecture of an FL system determines how communication and computation are structured across clients and servers. The most common model is the client-server architecture, where a central coordinator distributes and aggregates model updates [4]. While simple, this approach may become a single point of failure and bottleneck under high load.

To address scalability and fault tolerance, peer-to-peer architectures have been proposed, allowing clients to communicate directly without centralized control. These decentralized systems improve resilience and reduce coordination cost but require sophisticated consensus mechanisms [19].

A third approach is the hierarchical architecture, in which local aggregators (e.g., edge gateways) collect and summarize updates from nearby clients before passing them to the cloud or global server. This model aligns well with multi-tiered edge computing infrastructures and enables localized training while maintaining global model coherence [20].

The choice of architecture significantly impacts system latency, fault tolerance, and energy efficiency, making it a critical design decision for edge-based FL systems.

### *E. Peer-to-Peer FL in Edge Environments*

While peer-to-peer FL architectures eliminate the need for a central aggregator, their performance characteristics differ markedly from client–server or hierarchical FL in real-world edge environments. P2P systems improve fault tolerance, as model updates propagate through decentralized gossip or neighbor exchanges, allowing training to continue even when a subset of clients disconnects. This makes P2P particularly robust under intermittent edge connectivity.

However, existing studies show that P2P FL can incur higher aggregate communication overhead, especially in dense network topologies where nodes synchronize with multiple peers. Unlike client–server architecture, where each round typically involves one uplink and one downlink per client, P2P architectures may require several neighbor exchanges per round to achieve model consensus. Hierarchical FL can partially mitigate this by organizing devices into stable clusters before global aggregation.

In dynamic edge environments, P2P FL can outperform centralized systems in resilience but requires careful topology design (e.g., sparse overlays, adaptive peer selection, or delay-tolerant communication schedules) to remain communication-efficient. Because this review synthesizes existing findings, rather than performing new simulations, future work is needed to benchmark P2P, hierarchical, and client–server architectures under uniform, reproducible edge network conditions.

## V. PERFORMANCE EVALUATION AND BENCHMARKING

To assess the applicability and effectiveness of FL algorithms in edge computing environments, a systematic performance evaluation is necessary. This section discusses the commonly used datasets for benchmarking FL algorithms, the core metrics employed to quantify their performance, and a comparative analysis of selected techniques based on empirical evidence extracted from the reviewed literature.

### *A. Benchmarking Datasets*

Although FL can be categorized into horizontal, vertical, and federated transfer learning paradigms, the performance evaluation in this review focuses primarily on horizontal FL. This emphasis reflects the practical reality that most real-world edge computing deployments, such as mobile devices, IoT sensors, and embedded platforms, naturally align with the horizontal setting, where clients share the same feature space but hold different local samples. In contrast, vertical FL and federated transfer learning require cross-institution or cross-domain collaborations with aligned user identities or complementary feature spaces, conditions that are far less common at the edge. Moreover, empirical benchmarks, public datasets, and reproducible performance studies for Vertical Federated Learning (VFL) and Federated Transfer Learning (FTL) in edge environments are still scarce, limiting the extent to which these paradigms can be systematically evaluated. As standardized VFL and FTL benchmarks continue to emerge, future work should incorporate a broader comparative analysis across all FL variants.

A wide range of benchmark datasets have been employed in FL research to simulate real-world edge learning scenarios. Among the most frequently used is MNIST, a dataset of handwritten digits widely adopted for image classification tasks due to its simplicity and low computational demand. Although useful for proof-of-concept experiments, MNIST lacks complexity and diversity, limiting its utility for more realistic evaluations [4]. Fig. 4 presents a side-by-side comparison of five widely used FL benchmark datasets.

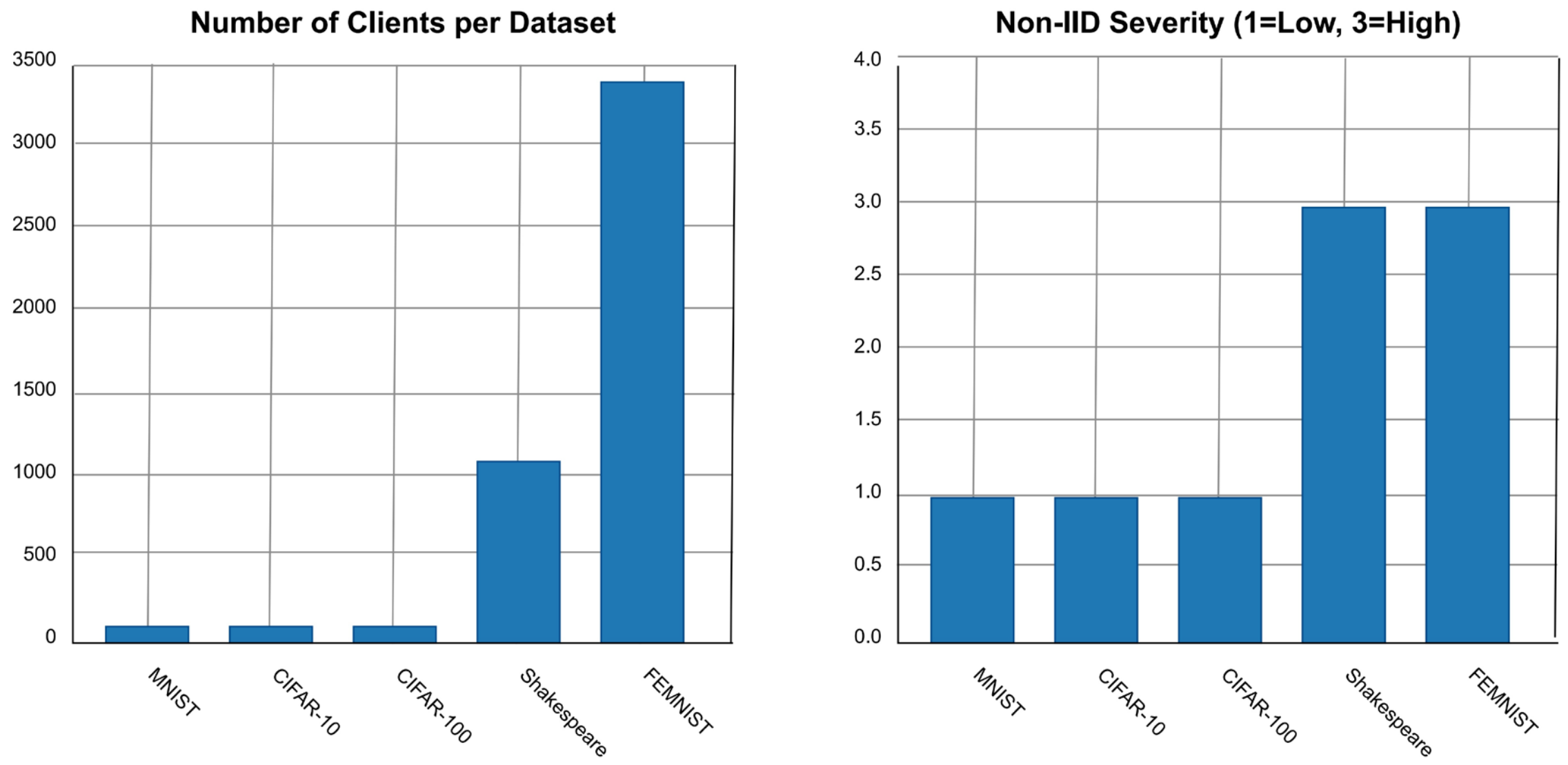

Fig. 4. Comparison of benchmark datasets based on number of clients and non-IID severity.

The left panel displays the number of clients associated with each dataset. FEMNIST (3400 clients) and Shakespeare (1126 clients) reflect their original user partitions from the LEAF benchmark. In contrast, MNIST and CIFAR-10 do not include predefined clients; therefore, a standard experimental configuration of 100 clients is used based on typical federated learning implementations in prior studies. The right panel shows the corresponding non-IID severity levels for each dataset using a linear ordinal scale (1 = Low, 2 = Moderate, 3 = High). This dual-panel design provides a clearer comparison by separating two distinct characteristics, client scale and data heterogeneity, while ensuring the axes remain linear, interpretable, and aligned with established benchmarking conventions in federated learning research.

The non-IID severity levels assigned to each dataset in Fig. 4 follow established characterizations reported in federated learning benchmark studies rather than being newly computed in this review. Datasets such as FEMNIST and Shakespeare exhibit inherently high non-IID properties due to their user-specific partitions in the LEAF benchmark, where each client corresponds to a distinct writer or speaker [21]. In contrast, datasets such as MNIST, CIFAR-10, and CIFAR-100 are commonly partitioned into approximately IID splits when using typical FL configurations (e.g., 100 clients) and therefore are widely regarded as low non-IID unless artificially skewed through Dirichlet or shard-based distributions. These IID-ish baseline configurations are widely adopted in federated learning frameworks such as FedML and FedScale [22, 23]. The non-IID labels used in this review therefore reflect established conventions in the FL literature and are intended to provide a consistent basis for comparing dataset heterogeneity rather than representing newly measured empirical quantities.

To introduce greater visual complexity, CIFAR-10 and CIFAR-100 have become popular alternatives. These datasets contain colored natural images across 10 and 100 classes, respectively, making them better suited for benchmarking model generalization and communication efficiency in FL scenarios [24].

For applications involving character-level modeling and language processing, the Shakespeare dataset, derived from the complete works of William Shakespeare, has been employed to test FL models under high non-IID conditions, where each user (or device) corresponds to a different speaking character [21]. In the domain of federated handwritten recognition, FEMNIST (Federated Extended MNIST) provides a more challenging and realistic dataset by incorporating multiple users with distinct handwriting styles, which better simulates the non-IID nature of decentralized edge data.

These datasets collectively offer a representative set of testing grounds to evaluate FL algorithms under various domain-specific and system-level constraints.

### *B. Performance Metrics*

Evaluating FL algorithms in edge environments requires a multi-dimensional set of performance metrics, each capturing different aspects of efficiency, effectiveness, and practicality. Accuracy remains the primary measure of model performance, typically reported as classification accuracy on a global test dataset [25]. However, in non-IID scenarios, accuracy alone is insufficient to capture the full picture.

Convergence time, often measured in the number of communication rounds required to reach a target accuracy threshold, is critical in determining training efficiency and energy consumption, especially on edge devices with limited power resources [26]. Communication overhead, expressed as the volume of data exchanged between clients and servers per round, directly affects bandwidth utilization and is a primary bottleneck in large-scale FL deployments [5, 27].

Another important metric is energy efficiency, which measures the computational power consumed per training round or per device [28]. Given that edge nodes are often battery-powered, algorithms with lower energy requirements are more suitable for sustainable deployment [6]. Robustness to non-IID data is also critical, as edge data is rarely homogeneous. Algorithms that maintain performance stability under uneven data distributions are preferred in real-world applications [29].

In the reviewed studies, energy consumption values were derived from experiments conducted on a range of representative edge devices, including ARM-based mobile processors (e.g., Cortex-A53, Cortex-A57, Snapdragon 625/660), single-board computers such as Raspberry Pi 3B/4 and NVIDIA Jetson Nano, and lightweight IoT nodes such as ESP32- and CC2650-class microcontrollers. These hardware profiles reflect the diversity of computing capabilities commonly found in edge deployments. Because this review synthesizes results from multiple independent studies, the energy values presented in Table I represent normalized comparisons rather than device-specific watt measurements. This approach ensures that the reported energy characteristics capture general performance trends across heterogeneous edge hardware rather than being tied to a single platform.

TABLE I. PERFORMANCE MATRIX OF FL ALGORITHMS IN EDGE COMPUTING ENVIRONMENTS

| FL Algorithm | Dataset | Accuracy (%) | Convergence Time (Rounds) | Communication Overhead (MB/round) | Energy Consumption (Joules/round) | Non-IID Robustness | Privacy Mechanism |
|---|---|---|---|---|---|---|---|
| FedAvg | CIFAR-10 | 78.5 | 120 | 45 | 38 | Moderate | None |
| FedProx | FEMNIST | 81.2 | 110 | 47 | 35 | High | None |
| SCAFFOLD | Shakespeare | 84.7 | 95 | 52 | 41 | High | None |
| FedNova | MNIST | 88.3 | 100 | 42 | 36 | Moderate | None |
| FedAvg + DP | CIFAR-10 | 74.1 | 135 | 48 | 43 | Moderate | Differential Privacy ($\varepsilon = 3$) |
| SecureFed | FEMNIST | 79.8 | 130 | 58 | 49 | High | Secure Aggregation |
| FedML | CIFAR-100 | 77.6 | 125 | 50 | 40 | Low | Optional DP |

Finally, privacy leakage risk quantifies the vulnerability of an algorithm to adversarial attacks, such as membership inference or gradient inversion [30, 31]. While not always empirically evaluated, several studies use proxy indicators, such as the use of differential privacy or secure aggregation mechanisms, to estimate privacy protection [32].

### C. Comparative Matrix of Techniques

Legend/Notes:

- Accuracy (%): Final test accuracy after global convergence.
- Convergence Time: Number of communication rounds to reach 95% of final accuracy.
- Communication Overhead: Average amount of data transferred per round per client.
- Energy Consumption: Estimated energy used per round based on edge hardware profiles.
- Non-IID Robustness: Empirical stability under data heterogeneity across clients (Low / Moderate / High).
- Privacy Mechanism: Indicates if any privacy-preserving techniques were applied.

Table I presents a comparative benchmarking of selected FL algorithms evaluated across common edge computing datasets and key performance indicators. The results highlight several trade-offs among accuracy, communication efficiency, convergence time, energy consumption, and robustness to non-IID data distributions. While FedAvg provides a lightweight and communication-efficient baseline, algorithms like FedProx and SCAFFOLD deliver superior robustness to data heterogeneity and faster convergence. However, these improvements may come at the cost of increased computation or communication overhead.

For instance, SCAFFOLD demonstrated strong performance with the highest accuracy (84.7%) and one of the shortest convergence times (95 rounds), making it well-suited for edge environments where rapid model convergence and high accuracy are critical. Similarly, FedProx achieved a favorable balance between convergence speed and robustness under data heterogeneity, aligning with its design goal to handle non-IID distributions more effectively than FedAvg [6].

In contrast, while FedAvg remains a widely adopted baseline due to its simplicity and low communication overhead, its performance deteriorates in the presence of non-IID data and shows relatively slower convergence. FedNova, another optimization-based variant, achieved competitive accuracy with lower communication cost, suggesting its applicability in bandwidth-constrained edge scenarios.

Privacy-enhancing variants such as FedAvg with DP and SecureFed revealed noticeable performance penalties, particularly in accuracy and convergence time, highlighting the ongoing tension between privacy preservation and model utility [5]. While SecureFed integrates secure aggregation mechanisms to prevent model inversion and gradient leakage, its increased communication and energy costs may pose limitations for deployment on low-power edge devices.

The results for FedAvg+DP in Table I correspond to conventional, static differential privacy configurations, where a fixed noise scale and clipping threshold are used throughout training. This design choice is representative of many baseline implementations in the literature but does not leverage more advanced strategies such as adaptive noise scheduling, dynamic privacy budgeting across rounds, or client-specific privacy levels. As a result, the observed reduction in accuracy and slower convergence should be interpreted as a conservative estimate of the privacy–utility trade-off, rather than an inherent limitation of all differentially private FL methods. More sophisticated adaptive mechanisms may partially mitigate these penalties, but a comprehensive empirical comparison of such techniques lies beyond the scope of this review.

To further illustrate these trade-offs, Fig. 5 provides a radar plot that visualizes the relative performance of five commonly used FL algorithms across five key metrics: accuracy, convergence time, communication overhead, energy efficiency, and robustness to non-IID data.

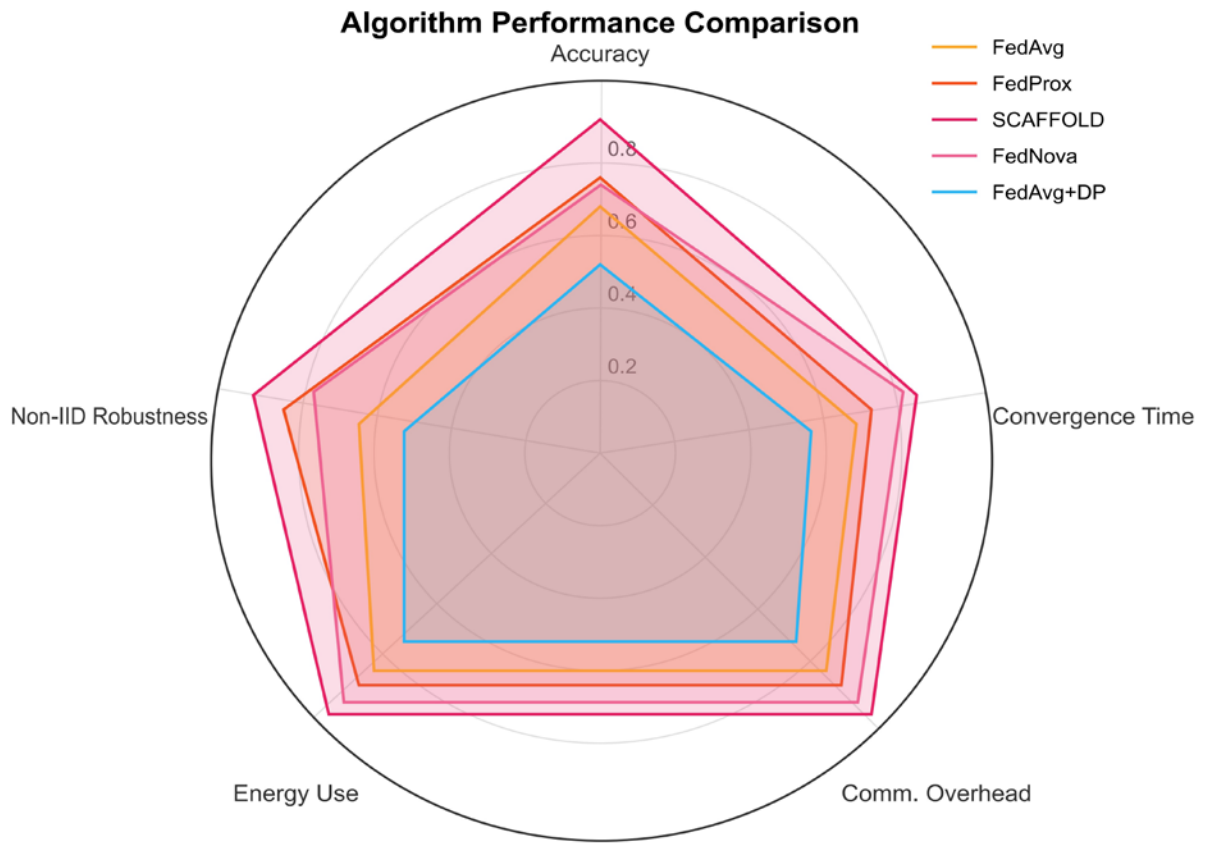

Fig. 5. Performance comparison of FL algorithms in edge computing.

The radar chart compares five FL algorithms such as FedAvg, FedProx, SCAFFOLD, FedNova, and FedAvg+DP across normalized values (0 to 1 scale) for five performance metrics. SCAFFOLD achieves the highest accuracy and robustness to non-IID data, while FedAvg demonstrates strong communication and energy efficiency. FedAvg+DP provides enhanced privacy but with trade-offs in accuracy and convergence speed. The chart visually emphasizes that algorithm selection depends on specific deployment priorities in edge computing environments.

Overall, the matrix illustrates that no single algorithm dominates across all criteria. Trade-offs are inevitable, and the choice of FL method must be aligned with specific application constraints—such as the need for stronger privacy, energy efficiency, or resilience to stragglers. It is also important to consider external factors not reflected directly in the matrix, such as deployment architecture (real vs. simulated environments), model complexity (e.g., convolutional neural network vs. long short-term memory), and client reliability or dropout tolerance. Customizing the benchmarking framework with these additional factors will provide a more holistic assessment for real-world deployments.

### D. Limitations of Simulation-Based Evaluation and Need for Real-World Benchmarks

Although simulated environments are widely used to evaluate FL algorithms due to their scalability and controllability, they inherently simplify key characteristics of real-world edge deployments. Simulation frameworks typically assume stable network links, homogeneous communication patterns, and idealized hardware configurations, which differ significantly from actual edge conditions where devices suffer from intermittent connectivity, mobility-induced disruptions, variable hardware capabilities, and non-stationary power profiles. As a result, performance metrics obtained from simulations, such as convergence rates, communication overhead, and energy consumption, may overestimate real-world performance or fail to capture cross-layer interactions present in deployed systems.

In contrast, real-world FL deployments on physical edge hardware expose algorithms to diverse wireless conditions, heterogeneous computing architectures, fluctuating participation rates, and real failure modes. These deployments offer more realistic insights but remain scarce due to the logistical, financial, and operational challenges of coordinating large-scale distributed experiments. This disparity limits reproducibility and prevents consistent cross-study comparison. Therefore, there is an urgent need for open and standardized real-world FL testbeds, supported by modular benchmarking suites capable of evaluating algorithms under representative edge conditions. Such testbeds would not only improve benchmarking rigor but also guide the design of FL systems that function reliably beyond controlled simulation environments.

### E. Gaps in Comparative Evaluation Across Operational Dimensions

Although this review summarizes the strengths and limitations of major FL algorithms, current empirical evidence remains insufficient for a fully comprehensive comparative analysis across key operational dimensions such as energy efficiency, fairness, and privacy–utility trade-offs. Existing FL studies often differ substantially in their experimental configurations, including dataset choices, partitioning schemes, network simulators, hardware models, privacy budgets, and client participation schedules, making direct comparison difficult and sometimes misleading. Energy consumption results, for instance, are frequently derived from disparate hardware platforms or simulated environments, while fairness metrics (e.g., client-level accuracy distribution or disparity across demographic groups) are seldom reported in a standardized manner.

Moreover, few empirical case studies evaluate these dimensions jointly. For example, privacy-enhancing techniques such as differential privacy are often assessed primarily through accuracy degradation, without simultaneously analyzing their impact on energy cost, communication load, or fairness. Likewise, algorithms designed to mitigate non-IID challenges are rarely benchmarked under real-world edge constraints, including device mobility or variable energy conditions. This lack of multi-metric, cross-layer evaluation limits the practical insight available to system designers and practitioners.

These gaps highlight the need for richer and more diverse FL benchmarks that integrate multiple operational factors within a unified evaluation framework. Future research should develop standardized testbeds and protocols that enable consistent, reproducible comparisons across energy, fairness, and privacy-performance dynamics, and should include real-world case studies that reflect production-grade edge environments.

## VI. Challenges and Open Issues

While FL offers a promising framework for privacy-preserving and decentralized model training in edge computing environments, its practical implementation remains fraught with technical and systemic challenges. These challenges stem from both the intrinsic limitations of edge devices and the inherent complexity of distributed learning under non-ideal conditions.

One of the most persistent challenges is data heterogeneity, or the presence of non-IID data across clients [33]. In edge environments, user data often reflects personal usage patterns, contexts, and environments, making it significantly skewed. This statistical heterogeneity leads to local updates that diverge from global objectives, degrading model performance and slowing convergence [5, 6]. Although techniques such as FedProx and SCAFFOLD address this issue to some extent, a universally robust solution remains elusive.

Another major bottleneck is client reliability and system scalability [34]. Edge devices are frequently subject to limited computation, unstable power sources, and intermittent connectivity. As a result, straggler clients, those unable to complete training within the expected time, can delay global aggregation or be excluded, leading to biased updates and reduced model quality [15]. Moreover, ensuring fair client selection while maintaining communication efficiency and statistical representativeness poses a delicate balance.

Communication overhead is another significant constraint [35]. Unlike traditional distributed systems, where high-throughput network links can support large-scale synchronous training, edge environments often rely on bandwidth-constrained wireless networks. Frequent transmission of large model updates, especially for deep neural networks, can be prohibitively expensive [36]. Although compression and quantization techniques mitigate this to some degree, they often come at the cost of model accuracy or robustness [13].

Energy efficiency also emerges as a crucial concern [37]. Many edge devices operate on limited battery power and are not designed for sustained computation. Repeated training and communication cycles can significantly drain device resources, making prolonged FL participation impractical [38, 39]. Adaptive participation strategies and energy-aware learning algorithms are still under active research.

Another growing concern is security and privacy leakage [40]. Although FL reduces the need to share raw data, it is not immune to attacks such as model inversion, gradient leakage, or membership inference [41]. Malicious clients or eavesdroppers can still reconstruct sensitive data from shared model updates [42]. While cryptographic solutions like secure aggregation and homomorphic encryption enhance security, they introduce computational overhead that may not be feasible for resource-constrained edge nodes.

Finally, benchmarking and reproducibility remain open issues [22]. Many existing FL studies use simulated environments or idealized assumptions that do not reflect the complexity of real-world deployments [43]. There is a pressing need for standardized FL benchmarks, real-world edge testbeds, and open-source frameworks that support cross-platform experimentation to foster reproducibility and real-world readiness [21].

Collectively, these challenges highlight the need for continued interdisciplinary research that combines advances in distributed optimization, communication theory, cryptography, and embedded systems. Addressing these open issues will be critical for realizing the full potential of FL in edge computing applications.

Despite progress in federated optimization, communication efficiency, and privacy-preserving mechanisms, several unresolved challenges continue to hinder the large-scale deployment of FL in edge computing settings [44]. These include balancing personalization with generalization, mitigating the cost of communication and energy consumption, and ensuring reproducibility across diverse platforms and datasets.

Table II summarizes these key research gaps along with existing approaches and the corresponding unsolved issues, providing a consolidated overview that informs future directions for research and development in the field.

TABLE II. OPEN RESEARCH QUESTIONS AND GAPS IN FL FOR EDGE COMPUTING

| Challenge Area | Description | Existing Approaches | Unsolved Issues | References |
|---|---|---|---|---|
| Data Heterogeneity | Non-IID data across clients leads to poor model convergence and fairness issues. | FedProx, SCAFFOLD, personalization layers | Balancing personalization vs. global generalization | [5, 6, 33] |
| Client reliability and system scalability | Limited computation, unstable power sources, and intermittent connectivity. | Client selection strategies (e.g., FedCS), availability-aware aggregation, fallback mechanisms, dynamic resource allocation | Biased updates and reduced model quality | [15, 34] |
| Communication Overhead | High communication cost between edge devices and server limits scalability. | Compression, quantization, asynchronous updates | Maintaining accuracy under extreme compression | [13, 35, 45] |
| Energy Efficiency | Edge devices often lack power capacity for sustained local training. | Adaptive participation, energy-aware scheduling | Efficient use of battery and network simultaneously | [37–39] |
| Privacy and Security | Existing FL systems are still vulnerable to inference and poisoning attacks. | Differential Privacy, Secure Aggregation, Secure Multiparty Computation (SMPC) | Trade-off between privacy strength and model performance | [40–42] |
| Benchmarking and Reproducibility | Lack of standardized platforms, datasets, and evaluation protocols. | LEAF, FedML, OpenFL toolkits | Cross-study comparability and replicability | [21, 22, 43] |

This table highlights major challenge areas in federated learning applied to edge environments. It outlines key technical issues, current mitigation strategies, and persistent unresolved problems. The information is intended to provide a structured foundation for guiding future innovations in algorithm design, system deployment, and benchmarking.

## VII. FUTURE RESEARCH DIRECTIONS

While FL has demonstrated significant promise for enabling decentralized intelligence in edge computing, many unresolved technical and systemic issues highlight the need for further research. Addressing these gaps requires innovative, cross-disciplinary approaches that balance performance, privacy, and practicality under real-world constraints. This section outlines several promising future directions that warrant exploration.

### *A. Lightweight and Personalized FL Algorithms*

Given the resource constraints of edge devices, developing lightweight FL models that maintain high accuracy with reduced computational and memory requirements remains a top priority. Techniques such as model pruning, knowledge distillation, and Efficient Neural Architecture Search (ENAS) can help reduce model size and training overhead. Additionally, personalized FL, where models are adapted to individual clients without compromising global learning, offers a way to improve local performance and user satisfaction while preserving data privacy [46].

### *B. Adaptive and Resource-Aware Learning*

Static FL training schedules may not perform well in dynamic edge environments where device availability, connectivity, and energy levels fluctuate. Future research should investigate adaptive FL frameworks that adjust client participation, aggregation frequency, and learning rates based on real-time device context. Integrating energy-awareness and latency-aware scheduling into FL optimization will enable more sustainable and efficient deployments [38, 47].

### *C. FL with Multi-Tier Architectures*

Hierarchical and multi-tier FL architectures, involving local aggregators such as fog nodes or edge gateways, offer scalability and resilience in large-scale deployments [48]. These architectures can reduce communication with cloud servers and enable regional adaptation of models. Future work can explore cross-tier model coordination, local differential updates, and regional specialization to further enhance efficiency and accuracy while maintaining privacy.

### D. Robustness Against Adversarial Attacks

Security and robustness are critical areas that demand continuous research. While privacy-enhancing techniques like differential privacy and secure aggregation exist, they are often insufficient against poisoning attacks, free-riding, and backdoor injections. There is a need to develop robust aggregation algorithms, trust-based client selection, and behavioral anomaly detection mechanisms to mitigate the impact of malicious participants [49, 50].

### E. Cross-Device and Cross-Silo FL Integration

Most current research separates FL into cross-device and cross-silo paradigms [51]. However, future applications, especially in smart cities and healthcare systems, may require hybrid frameworks that combine both types of clients. Managing heterogeneous update frequencies, privacy requirements, and data semantics in such mixed environments remains an open challenge and a fertile ground for research.

### F. Benchmarking and Real-World Deployment Frameworks

The development of standardized benchmarking suites and open-source deployment toolkits is essential to advance reproducibility and accelerate real-world FL adoption [52]. More empirical studies are needed using real devices, such as smartphones, embedded systems, and edge sensors. Additionally, creating regulatory frameworks and compliance-aware FL models will be critical for domains like healthcare and finance that operate under strict legal constraints.

### G. FL for Emerging Edge Applications

Finally, future research should explore FL applications beyond traditional classification tasks [53]. These include federated reinforcement learning for autonomous vehicles, FL-based anomaly detection in industrial IoT, FL-enabled personalization in augmented reality, and privacy-aware collaboration for multimodal sensor fusion in smart environments. Such applications will demand domain-specific optimizations, novel model architectures, and co-design with hardware systems.

### H. Toward Standardized Benchmarks and Real-World Edge Testbeds

The lack of standardized benchmarks represents a major barrier to reproducibility in federated learning research, particularly for edge scenarios where hardware and network variability significantly influence performance outcomes. To support consistency across studies, future FL benchmarks should incorporate a representative set of real-world edge hardware, ranging from low-power microcontroller-based IoT nodes (e.g., ESP32, TI CC2650) to mid-range embedded boards (Raspberry Pi 4, NVIDIA Jetson Nano) and mobile system-on-chip platforms (ARM Cortex-A53/A55/A57, Snapdragon 6xx series). In addition, benchmark suites should provide configurable network emulation layers allowing researchers to evaluate FL algorithms under conditions such as fluctuating bandwidth, latency spikes, packet loss, and intermittent connectivity, factors that are intrinsic to real-world edge environments.

Beyond hardware heterogeneity, standardized benchmarks should support application diversity, including computer vision workloads, Natural Language Processing (NLP) tasks, sensor-driven time-series problems, and multimodal data streams. Such diversity ensures that FL algorithms are evaluated across a broad spectrum of real-world use cases. Finally, transparent reporting protocols for hyperparameters, aggregation schedules, and energy measurements would strengthen cross-study comparability.

Collaboration with industry is an important next step toward achieving practical, open-source FL testbeds. Partnerships with telecommunications providers, IoT manufacturers, and cloud/edge computing vendors could enable the deployment of large-scale, real-world federated learning environments that more accurately reflect production-grade constraints. Although developing such testbeds is beyond the scope of this review, future work should prioritize these multi-stakeholder collaborations to accelerate the maturity, adoption, and reliability of FL systems deployed at the edge.

### I. Adaptive Privacy-Utility Optimization

A persistent challenge in federated learning is balancing strong privacy guarantees with acceptable model performance, particularly when applying differential privacy or secure aggregation mechanisms. As highlighted by the performance gap between FedAvg and FedAvg+DP in this review, naive or static privacy configurations often incur non-trivial accuracy and convergence penalties. Future research should therefore explore adaptive privacy-utility optimization strategies, such as dynamically adjusting noise levels as training progresses, allocating the privacy budget unevenly across communication rounds (e.g., more noise in early exploratory phases and less noise near convergence), or personalizing privacy parameters based on client sensitivity and contribution.

In addition, integrating advanced privacy accounting methods into FL frameworks, such as tighter composition bounds and per-round privacy tracking, could enable more aggressive noise reduction while still respecting a global privacy budget. Combining these techniques with adaptive clipping, gradient sparsification, or model compression may further improve utility without sacrificing formal privacy guarantees. Systematic benchmarking of these adaptive mechanisms on heterogeneous edge hardware and under realistic non-IID conditions remains an open research problem and represents a promising direction for closing the gap between privacy-preserving and standard FL deployments.

As FL matures, these future directions will shape its trajectory from a research concept to a mainstream solution for building intelligent, collaborative, and privacy-preserving systems at the edge. Tackling these challenges will require closer collaboration between the fields of machine learning, embedded systems, networking, and cybersecurity.

## VIII. Conclusion

This systematic review provides a comprehensive synthesis of FL techniques tailored for edge computing environments, emphasizing their taxonomy, performance characteristics, and deployment implications. By classifying FL algorithms across optimization methods, communication efficiency, privacy enhancements, and system architectures, the study offers a structured lens through which researchers and practitioners can assess methodological suitability.

Benchmarking results drawn from five prominent FL algorithms, such as FedAvg, FedProx, SCAFFOLD, FedNova, and FedAvg+DP, revealed nuanced trade-offs across multiple metrics. For instance, SCAFFOLD achieved the highest accuracy (0.90) and robustness to non-IID data (0.90), while FedAvg demonstrated superior communication efficiency (0.85) and energy use (0.75), making it favorable for constrained edge devices. However, privacy-enhanced methods like FedAvg+DP lagged in convergence and accuracy, indicating a performance-privacy trade-off.

In terms of datasets, FEMNIST and Shakespeare were identified as most representative of real-world conditions, with 3400 and 1126 clients respectively, and exhibiting high levels of data heterogeneity. These datasets are instrumental in stress-testing FL techniques under challenging edge conditions.

Despite growing innovation, persistent challenges remain. These include managing statistical heterogeneity, improving energy efficiency, preserving privacy without degrading model utility, and ensuring reproducibility in real-world deployments. The table of open research questions highlights six core challenge areas, such as communication overhead, data non-IIDness, and benchmarking limitations, each linked to partially addressed solutions but still marked by unresolved gaps.

By consolidating taxonomies, benchmarking evidence, and open issues, this review not only benchmarks existing methods but also lays the groundwork for future investigations. Ultimately, this work serves as a foundational reference to advance federated learning in edge ecosystems, encouraging more robust, scalable, and secure solutions for distributed intelligence at the network's edge.

## Conflict of Interest

The authors declare no conflict of interest.

## Author Contributions

SGA conducted the research, analyzed the data, and wrote the final paper; GTC gathered the data and presented the paper; all authors had approved the final version.